\documentclass[twocolumn,           
               showpacs,            
               nopreprintnumbers,     
               aps,                 
               prd,          	    
               letterpaper,             
              groupeaddress,      
               nofootinbib,         
               tightenlines,        
               floats,floatfix,      
               showkeys
               ]{revtex4-1}
               
\usepackage[toc,page]{appendix}
\usepackage{graphicx}
\usepackage{dcolumn}
\usepackage{bm}
\usepackage{amsmath}
\usepackage{amsfonts,amssymb}
\usepackage{soul}
\usepackage{color}
\definecolor{v}{rgb}{0.6, 0.2, 0.8} 
\definecolor{MAGA}{rgb}{0.1, 0.43, 0.75}
\definecolor{jm}{rgb}{0.13, 0.48, 0.64}

\usepackage{xcolor}
\usepackage{enumerate}
\usepackage{float}
\usepackage{subfigure}
\usepackage{ulem}
\newcommand\mathplus{+}

\usepackage{tabularx}

\begin{document}

\title{Constraining über gravity with recent observations and studying the $H_0$ problem}

\author{Gustavo A. Concha Valdez$^{1}$}
\email{gustavo.concha1@unmsm.edu.pe}
\author{Claudia Quintanilla$^{2}$}
\email{cquintanillach@unsa.edu.pe}
\author{Miguel A. Garc\'ia-Aspeitia$^{3}$}
\email{angel.garcia@ibero.mx}
\author{A.  Hern\'andez-Almada$^4$}
\email{ahalmada@uaq.mx}
\author{V. Motta$^5$}
\email{veronica.motta@uv.cl}

\affiliation{$^1$Grupo de Física Teórica, Facultad de Ciencias Físicas, Universidad Nacional Mayor de San Marcos, Ciudad Universitaria, Cercado de Lima 15081, Perú.}
\affiliation{$^2$Facultad de Ciencias Naturales, Universidad Nacional de San Agustín, Arequipa 04000, Perú.}
\affiliation{$^3$Depto. de Física y Matemáticas, Universidad Iberoamericana Ciudad de México, Prolongación Paseo \\ de la Reforma 880, México D. F. 01219, México.}
\affiliation{$^4$Facultad de Ingenier\'ia, Universidad Aut\'onoma de 
Quer\'etaro, Centro Universitario Cerro de las Campanas, 76010, Santiago de
Quer\'etaro, M\'exico,}
\affiliation{$^5$Instituto de F\'isica y Astronom\'ia, Universidad de Valpara\'iso, Avda. Gran Breta\~na 1111, Valpara\'iso, Chile.}

\begin{abstract}
This paper studies both $\Lambda$CDM and CDM models under the \"uber gravity theory, named  \"u$\Lambda$CDM and \"uCDM respectively. We report bounds over their parameter phase-space using several cosmological data, in particular, the recent Pantheon+ sample. Based on the joint analysis, the best fit value of the \"uber characteristic parameter is $z_\oplus = 0.046^{+0.047}_{-0.032}$ and  $z_\oplus = 1.382^{+0.020}_{-0.021}$ at 68\% confidence level for \"u$\Lambda$CDM and \"uCDM respectively. Although \"uber gravity can successfully mimics the cosmological constant, we find that the $\mathbb{H}0(z)$ diagnostic suggests the $H_0$ tension is not alleviated.
Finally, both models are statistically compared with $\Lambda$CDM through the Akaike and Bayesian information criteria. Both \"uber gravity models and $\Lambda$CDM are equally preferred for most of the single samples, in particular, \"u$\Lambda$CDM is not rejected by the CMB data. However, there is strong evidence against them for the joint analysis.
\end{abstract}
\pacs{}
\maketitle

\section{Introduction}

The $\Lambda$ Cold Dark Matter ($\Lambda$CDM) model is one of the most successful theories that explain with great precision the beginning and the evolution of our Universe at its current stages. However, among its many conundrums, the dark matter and energy components are the most intriguing due to the extremely weak interactions (or possible negligible) with the standard fields, being only the gravitational field the sole probe of its existence. The dark energy (DE) is related to the current Universe acceleration first observed by the supernova teams lead by Riess, Perlmutter among others \cite{Riess:1998,Perlmutter:1999} and later sustained by the Planck Satellite \cite{Planck:2018}. Clearly an acceleration of the space-time is not expected, thus it requires an unnatural dynamics: the addition of an extra component into the Friedmann equations which is able to create an accelerated expansion of the Universe and produce a late de Sitter scale factor. Without doubts, the cosmological constant is the cheapest way to tackle this problem, being an essential part of the $\Lambda$CDM paradigm, having also an exquisite capability to simulate the late Universe \cite{Carroll:2000}. However, it is important to emphasize that the cosmological constant is afflicted with several problems like those related to the vacuum energy density and the coincidence problem \cite{Zeldovich,Weinberg}.
Is in this vein that the community is exploring diverse alternatives to the cosmological constant, being the problem not yet settled (see \cite{Motta:2021hvl} for a compilation). 

Additionally, cosmologists concentrate most of their efforts in tackling the case of the current expansion rate of the Universe, the so-called $H_0$ {\it tension}. The reduction of uncertainties has led to significant discrepancies between the Planck \cite{Planck:2018} and SH0ES \cite{Riess:2019cxk} collaborations, whose disagreements are approximately at $\sim5.0\sigma$. Whether these discrepancies are due to systematic errors or mistaken theoretical assumptions, the $H_0$ tension could represent a window into new physics. Although modifying the standard cosmological model, in order to address the tension, without compromising its success in other areas has proven to be a very difficult task \cite{Abdalla:2022yfr}, the model-dependence of early $H_0$ measurements may suggest that it is worth exploring models beyond $\Lambda$CDM. The proposals include early dark energy \cite{Karwal:2016vyq,Poulin:2018cxd,PhysRevD.101.063523}, extra-relativistic degrees of freedom \cite{Aloni:2021eaq, Aboubrahim:2022gjb,Anchordoqui:2011nh}, new early dark energy \cite{PhysRevD.103.L041303,Niedermann:2020dwg}, diffusion models \cite{Haba_2016,Koutsoumbas_2018}, holographic dark energy \cite{LI20041,Huang_2005,Zhang2014}, to mention some of them.

In the present study we focus on two proposals: the ü$\Lambda$CDM and üCDM models, based on the über-gravitational model \cite{Khosravi:2017aqq}, which understands the $H_0$ tension as a transition in the context of a gravitational theory \cite{Khosravi:2017hfi}. Über-gravity sets the idea of taking an ensemble average over all possible $f(R)$ theories \cite{Khosravi:2016kfb}, a process inspired by statistical mechanics. Although the criteria for assigning probabilities to each theory are not clear, the general properties of ü$\Lambda$CDM  and üCDM can be studied and it shows promise in alleviating the $H_0$ tension. Indeed, the model contains two branches where the Hubble parameter evolves in different ways, with a threshold region determined by the free parameter $z_{\oplus}$, which is the redshift of transition to the über gravity. In this sense, it is also possible to determine the deceleration and jerk parameters in order to elucidate the characteristics of the component responsible for the acceleration. As stated, ü$\Lambda$CDM contains an extra parameter in comparison with üCDM, which contains the same number as $\Lambda$CDM. 
Therefore, in the case of ü$\Lambda$CDM, besides introducing a statistical penalization (e.g. Akaike  (AIC) or Bayesian information criteria (BIC) respectively) due to the extra parameter, it also keep the unsolved problem associated to the cosmological constant.

Additionally, we constrain the free parameters through a Bayesian analysis, using recent observations of Observational Hubble Parameters (CMB), Type Ia Supernovae (SnIa), Baryon Acoustic Oscillations (BAO), HII Galaxies (HIIG), Cosmic Microwave Background radiation (CMB) and a joint analysis. Moreover, we implement a $H_0$ diagnostic technique to study a possible alleviation to the $H_0$ tension using the über-gravity paradigm.

The paper is organized as follows: in Sec. \ref{MB} we review the mathematical background of ü$\Lambda$CDM and üCDM models, emphasizing its main properties and defining the parameters to be analyzed. In Sec. \ref{sec:constraints} the data and methodology are described, and the results are shown in Sec. \ref{RE}. Finally, in \ref{CO} we summarize the conclusions and give some outlooks on the subject. We henceforth use units in which $\hbar=k_B=c=1$.

\section{Mathematical Background} \label{MB}

The idea of taking an ensemble average over gravity theories is condensed in the following Lagrangian
\begin{equation}
    \mathcal{L}=\left(\sum_{i=1}^N\mathcal{L}_ie^{-\beta\mathcal{L}_i}\right)\left(\sum_{i=1}^N e^{-\beta\mathcal{L}_i}\right)^{-1},
\end{equation}
where the index $i$ corresponds to the $i$-th theory considered through the Lagrangian associated $\mathcal{L}$, being N the dimension of the gravitational models space over the manifold $\mathbb{M}$. The free parameter $\beta$ assigns probabilities to each model. For greater rigor, the ensemble average is taken over all analytic models of gravity, described by the $f(R)$ family, where the $R$ stands for the Ricci scalar. However, in \cite{Khosravi:2017hfi} is shown that the Lagrangian, after a change of basis, can be written as 

\begin{equation}
    \mathcal{L}_{\mathrm{uber}}=\left(\sum_{n=1}^\infty(\Bar{R}^n-2\Lambda)e^{-\beta(\Bar{R}^n-2\Lambda)}\right)\left(\sum_{n=1}^\infty e^{-\beta(\Bar{R}^n-2\Lambda)}\right)^{-1},
\end{equation}
where $\Bar{R}\equiv R/R_0$ contains a new cosmological parameter $R_0$, $n$ is an integer, and $\Lambda$ is the cosmological constant. Next, we will show two models associated with ü$\Lambda$CDM and üCDM.

\subsection{ü$\Lambda$CDM model}

In background cosmology it is possible to assume  homogenity and isotropy in the line element through the Friedmann-Lemaitre-Robertson-Walker (FLRW) equations as $ds^2=-dt^2+a(t)^2[dr^2+r^2d\Omega^2]$, where $d\Omega^2=d\theta^2+\sin^2\theta d\varphi^2$ and $a(t)$ is the scale factor. Thus, the Friedmann equation in this scenario can be written as \cite{Khosravi:2017aqq}
\begin{equation}
    E^2(z)_{\mathrm{\ddot{u}\Lambda CDM}}= \left\{ \begin{array}{cc}
             \Omega_{0m}(z+1)^3+\Omega_{0r}(z+1)^4+\Omega_{0\Lambda}, & z>z_\oplus \\
             \\ \frac{1}{2}\Bar{R}_0+(1-\frac{1}{2}\Bar{R}_0)(z+1)^4, &  z<z_\oplus \\
             \end{array}
   \right. \label{Euber}
\end{equation}
where $E(z)\equiv H(z)/H_0$, $H(z)$ is the Hubble parameter, $z$ is the redshift (i.e. $H_0=H(z=0)$), $\Omega_{0m}$ is the matter density parameter, $\Omega_{0r}$ is the radiation density parameter, both at $z=0$ and $\Bar{R}_0\equiv R_0/6H_0^2$. In this case, $\Omega_{0\Lambda}$ plays a role at $z>z_\oplus$, being $z_\oplus$ the region of transition to über gravity.

Using the continuity for $E(z)$ and $E'(z)$, where the prime denotes a derivative with respect to $z$, $\bar{R}_0$ takes the form
\begin{equation}
    \frac{\Bar{R}_0}{2}=\frac{(1-\Omega_{0r})(z_\oplus+1)^4+3\Omega_{0\Lambda}}{3+(z_\oplus+1)^4}.
\end{equation}
One of the magnitudes of interest for the present study is the deceleration parameter, written in terms of the redshift, which is given by
\begin{equation}
   q(z)_{\mathrm{\ddot{u}\Lambda CDM}}= \left\{ \begin{array}{cc}
             \frac{3\Omega_{0m}(z+1)^3+4\Omega_{0r}(z+1)^4}{2[\Omega_{0m}(z+1)^3+\Omega_{0r}(z+1)^4+\Omega_{0\Lambda}]}-1, & z>z_\oplus \\
             \\ \frac{2(1-\frac{1}{2}\Bar{R}_0)(z+1)^4}{\frac{1}{2}\Bar{R}_0+(1-\frac{1}{2}\Bar{R}_0)(z+1)^4}-1, &  z<z_\oplus \\
             \end{array}
   \right.
   \label{decelparam}
\end{equation}
Moreover, the jerk parameter is given by the expression 
\begin{equation}
    j(z)_{\mathrm{\ddot{u}\Lambda CDM}}= \left\{ \begin{array}{cc}
             \frac{\Omega_{0m}(z+1)^3+3\Omega_{0r}(z+1)^4+\Omega_{0\Lambda}}{\Omega_{0m}(z+1)^3+\Omega_{0r}(z+1)^4+\Omega_{0\Lambda}}, & z>z_\oplus \\
             \\ \frac{2(1-\frac{1}{2}\Bar{R}_0)(z+1)^4}{\frac{1}{2}\Bar{R}_0+(1-\frac{1}{2}\Bar{R}_0)(z+1)^4}+1, &  z<z_\oplus \\
             \end{array}
   \right.
\end{equation}
which behaves like the standard $\Lambda$CDM model in the region $z>z_\oplus$,  while for $z<z_\oplus$ the über gravity dominates its evolution.

Finally, the transition redshift $z_T$, which determines the beginning of the accelerated expansion of the universe $(q(z_T)=0)$, was calculated for both models. For $z_T<z_\oplus$ the transition redshift takes the form
\begin{equation}
    z_T=\sqrt[4]{\frac{3\Omega_{0\Lambda}+(1-\Omega_{0r})(z_\oplus+1)^4}{3(1-\Omega_{0\Lambda})+\Omega_{0r}(z_\oplus+1)^4}}-1,
\end{equation}
whereas for $z_T>z_\oplus$ the value comes from solving
\begin{equation}
    \Omega_{0m}(z_T+1)^3+2\Omega_{0r}(z_T+1)^4=2\Omega_{0\Lambda},
\end{equation}
which is obtained numerically. The transition in this model occurs for $z_T>z_\oplus$, so in this study only this last expression is taken into account.

\subsection{üCDM model}

In this case, we propose that the Universe acceleration is only driven by über contributions, thus we consider $\Omega_{\Lambda}=0$, reducing the free parameters to the same number as in the standard model and assuming FLRW cosmology. Therefore, we have 

\begin{equation}
    E^2(z)_{\mathrm{\ddot{u} CDM}}= \left\{ \begin{array}{cc}
             \Omega_{0m}(z+1)^3+\Omega_{0r}(z+1)^4, & z>z_\oplus \\
             \\ \frac{1}{2}\Bar{R}_0+(1-\frac{1}{2}\Bar{R}_0)(z+1)^4, &  z<z_\oplus \\
             \end{array}
   \right. \label{EuberSinLambda}
\end{equation}
being
\begin{equation}
    \frac{\Bar{R}_0}{2}=\frac{(1-\Omega_{0r})(z_\oplus+1)^4}{3+(z_\oplus+1)^4}, \;\; \Omega_{0m}=\frac{4(1-\Omega_{0r})(z_{\oplus}+1)}{3+(z_{\oplus}+1)^4}.
\end{equation}
For the deceleration parameter we have
\begin{equation}
    q(z)_{\mathrm{\ddot{u} CDM}}= \left\{ \begin{array}{cc}
    \frac{3\Omega_{0m}(z+1)^3+4\Omega_{0r}(z+1)^4}{2[\Omega_{0m}(z+1)^3+\Omega_{0r}(z+1)^4]}-1, & z>z_\oplus \\
             \\ \frac{2(1-\frac{1}{2}\Bar{R}_0)(z+1)^4}{\frac{1}{2}\Bar{R}_0+(1-\frac{1}{2}\Bar{R}_0)(z+1)^4}-1, &  z<z_\oplus \\
             \end{array}
   \right.
   \label{decelparamCDM}
\end{equation}
and the jerk parameter reads 
\begin{equation}
    j(z)_{\mathrm{\ddot{u} CDM}}= \left\{ \begin{array}{cc}
             \frac{\Omega_{0m}+3\Omega_{0r}(z+1)}{\Omega_{0m}+\Omega_{0r}(z+1)}, & z>z_\oplus \\
             \\ \frac{2(1-\frac{1}{2}\Bar{R}_0)(z+1)^4}{\frac{1}{2}\Bar{R}_0+(1-\frac{1}{2}\Bar{R}_0)(z+1)^4}+1, &  z<z_\oplus \\
             \end{array}
   \right.
\end{equation}
being the first region $z>z_{\oplus}$ (for $q$ and $j$) dominated by matter while the later one is controlled by the über effects that drive the acceleration.

Finally, in \"uCDM the transition is not allowed in the region $z_T>z_\oplus$, giving, as a result, the following analytic expression, valid for $z_T<z_\oplus$
\begin{equation}
    z_T=\sqrt[4]{\frac{(1-\Omega_{0r})(z_\oplus+1)^4}{3+\Omega_{0r}(z_\oplus+1)^4}}-1,
\end{equation}
Note that $z_T$ shows a dependence only on $z_\oplus$ and $\Omega_{0r}$.

\section{Data and methodology} \label{sec:constraints}

Both \"u$\Lambda$CDM and \"uCDM cosmologies are confronted using CC, HIIG, SnIa, BAO, CMB and joint data through a Bayesian Markov Chain Monte Carlo (MCMC) analysis. To bound their free parameter phase-space, ($h$, $\Omega_{0\Lambda}$, $z_\oplus$) for \"u$\Lambda$CDM and ($h$, $z_\oplus$) for \"uCDM,  we use the \texttt{emcee} package \cite{Foreman:2013} under Python language. We establish a configuration to achieve the convergence of the chains using the autocorrelation function, and generate a set of 3000 chains with 250 steps. Additionally, we use a Gaussian prior over $h$ as $h=0.7403\pm 0.0142$ \cite{Riess:2019cxk}  and $h=0.6766\pm 0.0042$ \cite{Planck:2018} as a consistency probe, flat priors over $\Omega_{0\Lambda}$ and $z_{\oplus}$ in the region [0,1] and [0,2] respectively. Thus, the $\chi^2$-function is given by
\begin{equation}\label{eq:chi2}
    \chi_{\rm Joint}^2 = \chi_{\rm CC}^2 + \chi_{\rm HIIG}^2 + \chi_{\rm SnIa}^2  + \chi_{\rm BAO}^2+\chi_{\rm CMB}^2 \,,
\end{equation}
where each term corresponds to the $\chi^2$ function per sample.

\subsection{Cosmic chronometers}

We use a sample of 31 measurements of the Hubble parameter using differential age method \cite{Moresco:2016mzx} (see also \cite{Magana:2017}). Due these points are cosmological model independent, they are useful to test alternative cosmologies to $\Lambda$CDM. The $\chi^2$ function can be built as
\begin{equation} \label{eq:chiOHD}
    \chi^2_{{\rm OHD}}=\sum_{i=1}^{31}\left(\frac{H_{th}(z_i)-H_{obs}(z_i)}{\sigma^i_{obs}}\right)^2,
\end{equation}
where $H_{th}(z_i)$ is the theoretical Hubble parameter using Eq. \eqref{Euber}, and $H_{obs}(z_i)\pm \sigma_{obs}^i$ is the observational counterpart with its uncertainty at the redshift $z_i$.

\subsection{Type Ia Supernovae (Pantheon$\mathplus$)}

Recently, a sample of 1701 measurements of the luminosity modulus coming from SNIa, namely Pantheon$\mathplus$ sample, is reported by \cite{Scolnic2018-qf, Brout_2022}. This represents the largest sample and covers a region $0.001<z<2.26$. Considering that this sample is extracted from 1550 distinct SNIa, we build the $\chi^2$ function as
\begin{equation}\label{eq:chi2SnIa}
    \chi_{\rm SnIa}^{2}=a +\log \left( \frac{e}{2\pi} \right)-\frac{b^{2}}{e},
\end{equation}
where
\begin{eqnarray}
    a &=& \Delta\boldsymbol{\tilde{\mu}}^{T}\cdot\mathbf{Cov_{P}^{-1}}\cdot\Delta\boldsymbol{\tilde{\mu}}, \nonumber\\
    b &=& \Delta\boldsymbol{\tilde{\mu}}^{T}\cdot\mathbf{Cov_{P}^{-1}}\cdot\Delta\mathbf{1}, \\
    e &=& \Delta\mathbf{1}^{T}\cdot\mathbf{Cov_{P}^{-1}}\cdot\Delta\mathbf{1}, \nonumber
\end{eqnarray}
and $\Delta\boldsymbol{\tilde{\mu}}$ is the vector of the difference between the theoretical distance modulus and the observed one, $\Delta\mathbf{1}=(1,1,\dots,1)^T$, $\mathbf{Cov_{P}}$ is the covariance matrix formed by adding the systematic and statistic uncertainties. The transpose of the vectors are denoted with the super-index $T$ on the previous expressions.

The theoretical counterpart of the distance modulus is estimated by
\begin{equation}
    m_{th}=\mathcal{M}+5\log_{10}\left[\frac{d_L(z)}{10\, {\rm pc}}\right],
\end{equation}
where $\mathcal{M}$ is a nuisance parameter which has been marginalized by Eq. \eqref{eq:chi2SnIa}. The luminosity distance, denoted as $d_L(z)$, is computed through
\begin{equation}\label{eq:dL}
    d_L(z)=(1+z)c\int_0^z\frac{dz^{\prime}}{H(z^{\prime})},
\end{equation}
where $c$ is the speed of light.

\subsection{Baryon Acoustic Oscillations}

The Baryon Acoustic Oscillations signature is an standard ruler useful to constrain cosmological model parameters and is the result of the interactions between baryons and photons in the recombination era. We use 6 correlated points obtained by \cite{Percival:2010,Blake:2011,Beutler:2011hx} and collected by \cite{Giostri:2012}. We confront them with the cosmological models by building the $\chi^2$ function as
\begin{equation}\label{eq:bao}
 \chi^2_{\rm BAO}= \vec{X}^T {\rm C}_{BAO}^{-1} \vec{X}
\end{equation}
where $\vec{X}^T$ is the residual vector between observational measurements and theoretical values of the ratio $d_A(z_*)/D_V(z)$,  where $d_A(z_d)$ is the comoving angular diameter distance at the the photon decoupling epoch ($z_*$) and the dilation scale is given by \cite{Wigglez:Eisenstein2005}
\begin{equation}
    D_V(z) = \left[ d_A^2(z) c z / H(z)\right]^{1/3}
\end{equation}
where $c$ is again the speed of light. For this work we use $z_* = 1089.80 \pm 0.21$ \cite{Planck:2018}.

\subsection{HII Galaxies}

A sample of 181 measurements coming from Hydrogen II galaxies (HIIG), with their luminosity dominated by young massive burst of star formation, is reported by \cite{GonzalezMoran2019, Gonzalez-Moran:2021drc}. This sample which covers a region $0.01<z<2.6$ and is useful to establish bounds over cosmological parameters due the correlation between the measured luminosity $L$ of the galaxies and the inferred velocity dispersion $\sigma$ of their ionized gas \cite{Chavez2012,Chavez2014,Terlevich2015,Chavez2016}. The $\chi^2$-function is built as
\begin{equation}\label{eq:chi2_HIIG}
    \chi^2_{{\rm HIIG}}=\sum_i^{181}\frac{[\mu_{th}(z_i, {\Theta})-\mu_{obs}(z_i)]^2}{\epsilon_i^2},
\end{equation}
where $\epsilon_i$ is the observational uncertainty measured at $z_i$ having $68\%$ of confidence level. Additionally, the observational distance modulus ($\mu_{obs}$) is expressed
\begin{equation}
    \mu_{obs} = 2.5(\alpha + \beta\log \sigma -\log f - 40.08)\,.
\end{equation}
Here, $\alpha$ and $\beta$ are the intercept and slope of the $L$-$\sigma$ relation and $f$ is the measured flux. 
The theoretical estimate is given as
\begin{equation}
    \mu_{th}(z, \Theta) = 5 \log_{10} \left [ \frac{d_L(z, \Theta)}{1\,{\rm Mpc}}\right] + 25,
\end{equation}
where $d_L$ is the luminosity distance measured in Mpc (see Eq. \eqref{eq:dL}).

\subsection{Cosmic Microwave Background Radiation}

The CMB temperature anisotropies are useful measurements  to establish constraints over cosmological parameters. A way to use them without performing a full perturbative analysis is to compress the full information into some parameters. Authors in \cite{Chen_2019} compress the CMB information from Planck 2018 Temperature Power Spectrum (TT), for high TE multipoles, polarization spectra EE modes + lowE data \cite{Planck:2018} in the acoustic scale $l_A$ which characterizes the CMB temperature in the transverse direction, the shift parameter $R$ which influences the CMB temperature along the line-of-sight direction, and the quantity $\Omega_{b0}h^2$ where $\Omega_{b0}$ is the density of baryons today ($z=0$). Thus, the figure-of-merit is built as
\begin{equation}
    \chi^2_{\rm CMB}=V_{\rm CMB}{\rm Cov}^{-1}_{\rm CMB}V_{\rm CMB}^T,
\end{equation}
where $V_{\rm CMB}$ is 
\begin{equation}
 V_{\rm CMB} =\left(
 \begin{array}{c}
 R^{th} - 1.7493\\
 l_A^{th} - 301.462 \\
 \Omega_{b}h^{2 th} - 0.02239
\end{array}\right),
\end{equation}
the superscripts $th$ refers to the theoretical estimates, and $\rm Cov_{CMB}^{-1}$ represents the inverse of
\begin{equation}
\rm{Cov_{CMB}} = 10^{-8}\left(
\begin{array}{ccc}
2162.25 & 19560.23 & -46.04   \\
19560.23 & 801025.00 & -456.45\\
-46.04 & -456.45 & 2.25
\end{array}\right),
\label{eq:invcovcmb}
\end{equation}
which is the  covariance matrix for $ V_{\rm CMB}$. The theoretical  counterparts are estimated by
\begin{equation}
    l_A^{th} = (1+z_*)\pi \frac{D_A(z_*)}{r_s(z_*)},
\end{equation}
and 
\begin{equation}
    R^{th}(z_*) = \frac{(1+z_*)D_A(z_*)\sqrt{\Omega_{0m} H^2_0}}{c},
\end{equation}
where $z_*$ is the redshift at the photon decoupling phase that takes the form
\begin{equation}
    z_*=10488[1+0.00124(\Omega_bh^2)^{-0.738}][1+g_1(\Omega_mh^2)^{g_2}],
\end{equation}
where
\begin{eqnarray}
    &&g_1=\frac{0.0738(\Omega_bh^2)^{-0.238}}{1+39.5(\Omega_bh^2)^{0.763}}, \\
    &&g_2=\frac{0.560}{1+21.1(\Omega_bh^2)^{1.81}},
\end{eqnarray}
and $D_A$ is the angular diameter distance for a flat geometry that reads as
\begin{equation}
    D_A(z)=\frac{c}{H_0(z+1)}\int_0^z\frac{dz'}{E(z')},
\end{equation}
while $r_s$ is the comoving sound horizon given by
\begin{equation}
    r_s(z)=\frac{c}{H_0}\int_0^{1/(z+1)}\frac{da}{a^2E(a)\sqrt{3\left(1+\frac{3\Omega_bh^2}{4\Omega_{\gamma}h^2}a\right)}},
\end{equation}
being $\Omega_b$ and $\Omega_{\gamma}$ the baryons and photons density parameters respectively \cite{Hu:1995en}.
\section{Results} \label{RE}

\begin{figure*}
    \centering
    \includegraphics[width=0.6\textwidth]{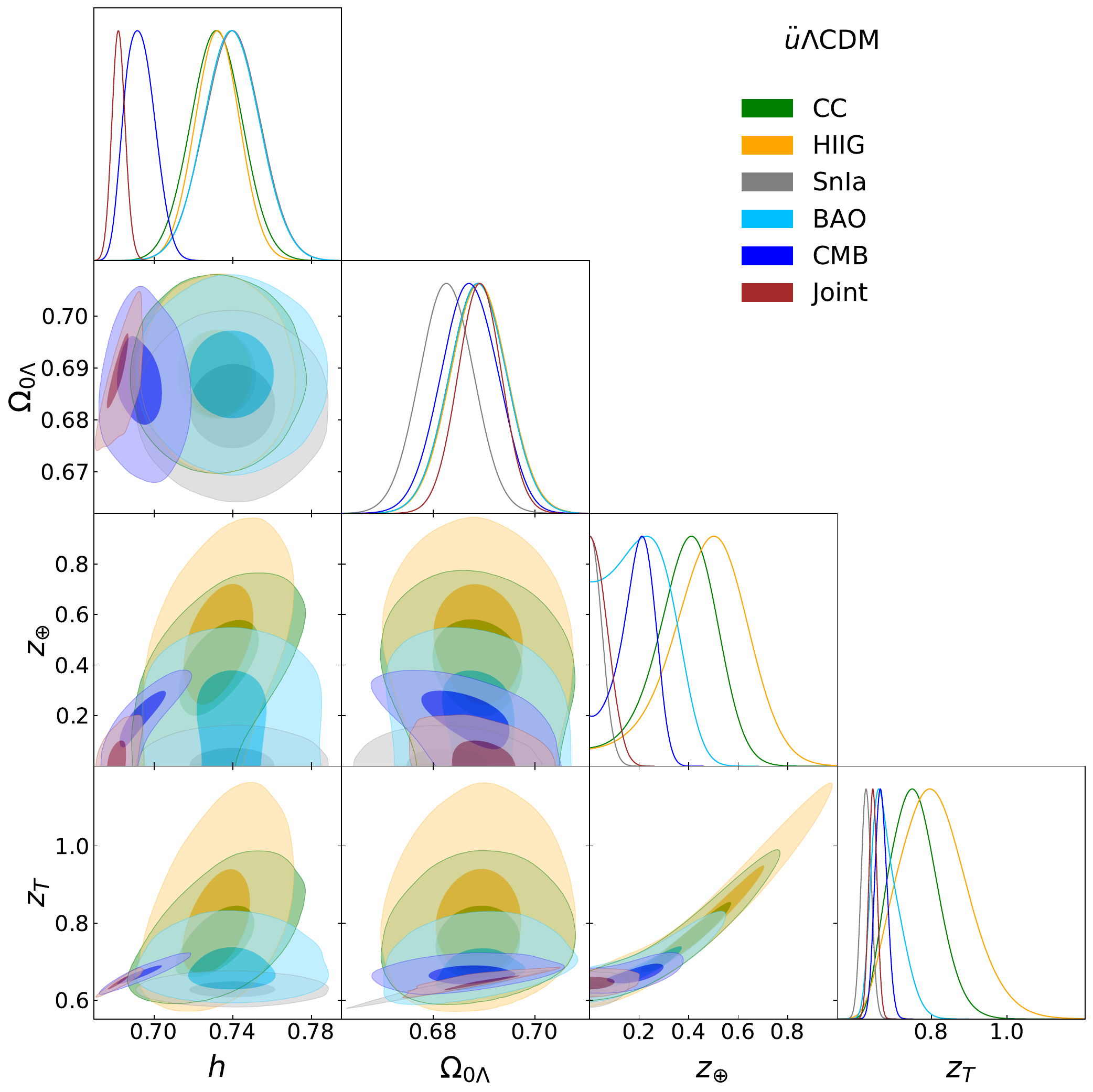}
    \caption{2D contours at $1\sigma$ (inner region) and $3\sigma$ (outermost region) CL for the ü$\Lambda$CDM.}
    \label{fig:contours}
\end{figure*}

\begin{figure*}
    \centering
    \includegraphics[width=0.5\textwidth]{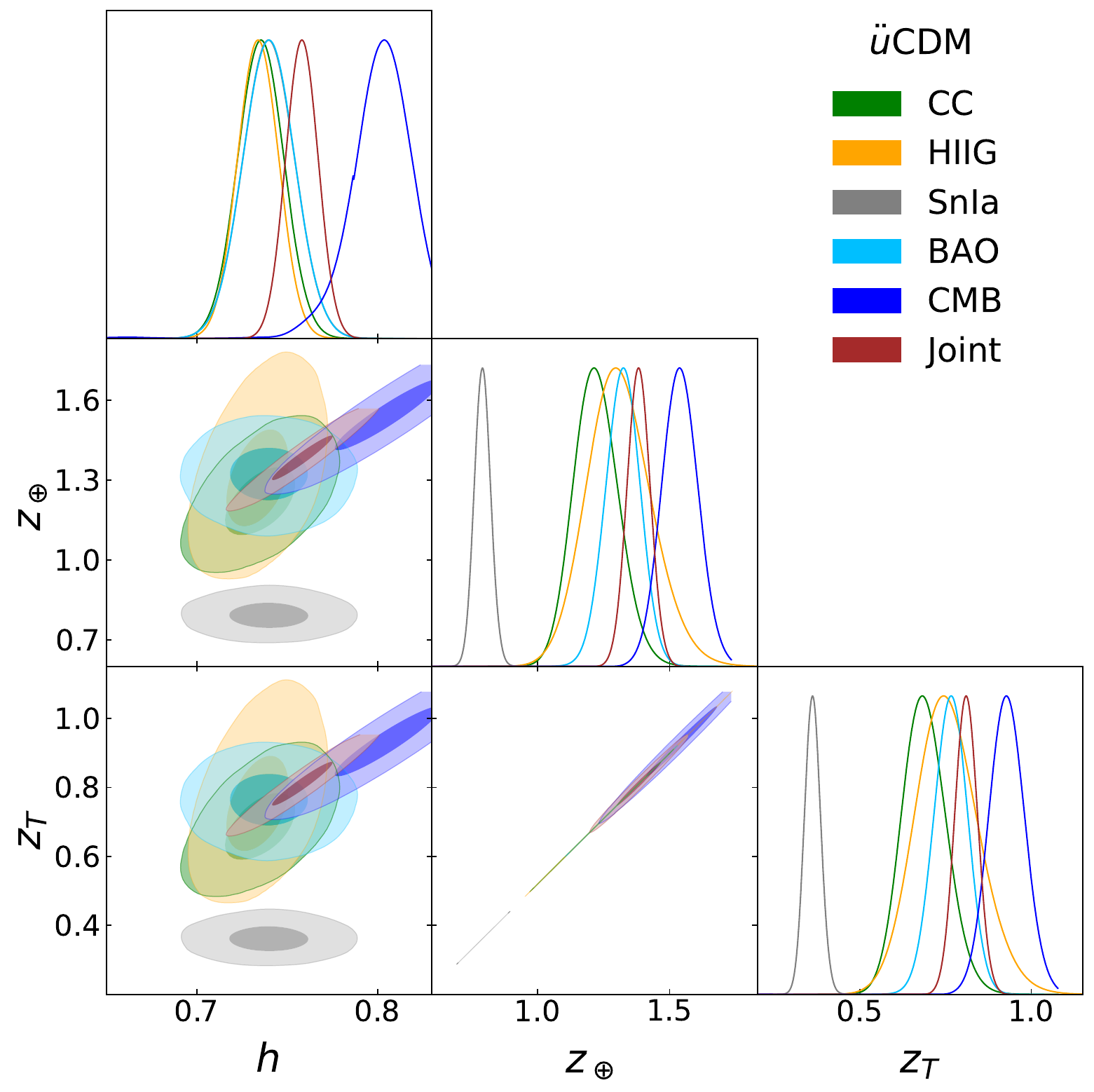}
    \caption{2D contours at $1\sigma$ (inner region) and $3\sigma$ (outermost region) CL for the üCDM.}
    \label{fig:contours1}
\end{figure*}

We use the über gravity in the context of ü$\Lambda$CDM and üCDM models to understand the nature of the current Universe acceleration and help us to interpret the origin of the $H_0$ tension. Due to the consistency between the results obtained by using $H_0$ Gaussian priors from Planck and SH0ES, hereafter we centered our discussion in the case of $H_0$ with priors coming from low-redshift (see Table \ref{tab:bestfits}). In Figs. \ref{fig:contours} and \ref{fig:contours1} we present the 2D parameter likelihood contours at $68\%$ (1$\sigma$) and $99.7\%$ (3$\sigma$) confidence level (CL) respectively for both models. Moreover Table \ref{tab:bestfits} shows the mean values of the parameters and their uncertainties at 1$\sigma$.

Additionally to this, we statistically compare these results with $\Lambda$CDM model, applying the corrected Akaike information criterion (AICc) \cite{Akaike:1974}
and the Bayesian information criterion (BIC) \cite{schwarz1978}. Both criteria penalize according to the size of the data sample ($N$) and the number of degrees of freedom ($k$) defined as ${\rm AICc}=\chi^2_{\rm min}+2k+2(k^2+k)/(N-k-1)$ and ${\rm BIC}=\chi^2_{\rm min}+k\log(N)$, respectively, where $\chi^2_{\rm min}$ is the minimum of $\chi^2$. Therefore, a model with a lower value of AICc (BIC) is preferred by the data. 

For the $\Delta {\rm AICc}$ we have the following conditions: if $|\Delta {\rm AICc}|<4$ both models are statistically equivalent, if $4<{\rm AICc}<10$ the data still support the given model (über gravity) but less than the preferred one (concordance model) and finally if ${\rm AICc}>10$ indicates that the data do not support the given model. In addition, for $\Delta {\rm BIC}$ the interpretation is as follows: if  $\Delta {\rm BIC}<2$ there is no evidence against the new model which in this case is über gravity, if $2<\Delta {\rm BIC}<6$ there is modest evidence against the new model, and finally if $\Delta {\rm BIC}>10$, gives the strongest evidence against it. 

We start by showing the AICc and BIC for the $\Lambda$CDM model: for AICc values we have 22.07 (CC), 440.72 (HIIG), 2002.41 (SnIa), 12.82 (BAO), 23.80 (CMB) and 2493.25 (Joint); as for BIC it has 24.51 (CC), 447.05 (HIIG), 2013.28 (SnIa), 6.04 (BAO), 21.56 (CMB) and 2504.37 (Joint). Table \ref{tab:bestfits} shows both $\Delta$AICc and $\Delta$BIC by considering $\Lambda$CDM as the reference scenario. According to AICc, we find that both \"u$\Lambda$CDM and $\Lambda$CDM are preferred equally by CC, HIIG and CMB  while the \"u$\Lambda$CDM is not supported by SnIa and BAO, but there is a still support for the \"uber gravity when the combined data are considered. Regarding BIC, we find the strongest evidence against \"u$\Lambda$CDM for SnIa and the combined data and a modest evidence against the uber gravity for HIIG.

As for \"uCDM, we observe that \"uCDM and $\Lambda$CDM are equally preferred by CC, HIIG and BAO while the strongest evidence against \"uCDM is provided by SnIa and Joint analysis, and a strong evidence against \"uber gravity is given by CMB. It is worth to mention that $\Delta$AICc and $\Delta$BIC are reduced to $\Delta \chi^2$ because both models have same number of free parameters.

\begin{figure*}
\centering
\includegraphics[width=0.32\textwidth]{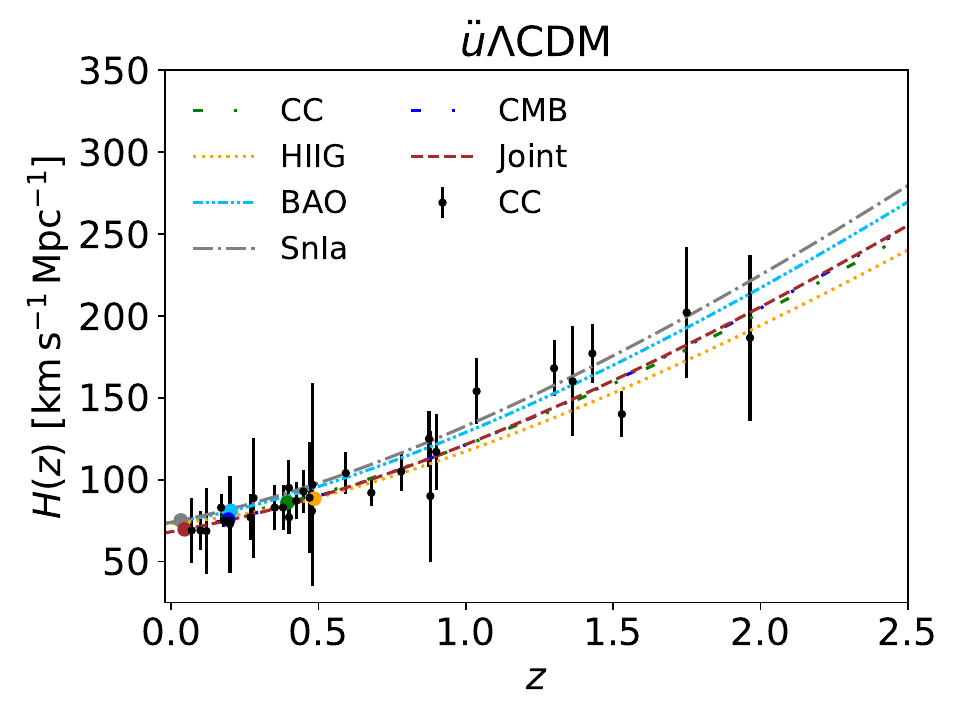}
\includegraphics[width=0.32\textwidth]{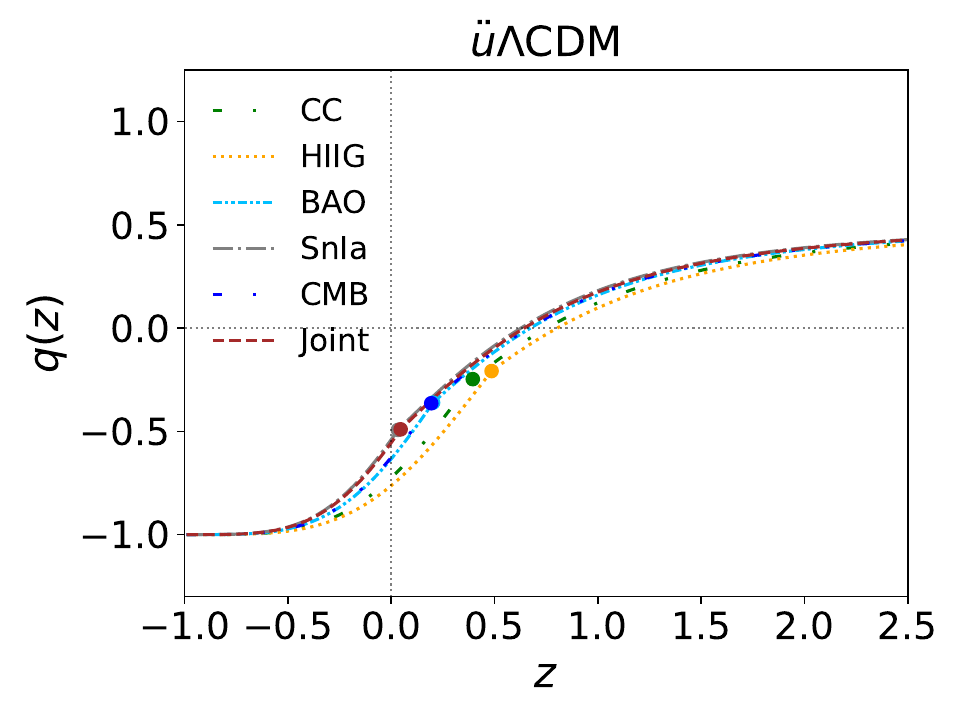}
\includegraphics[width=0.32\textwidth]{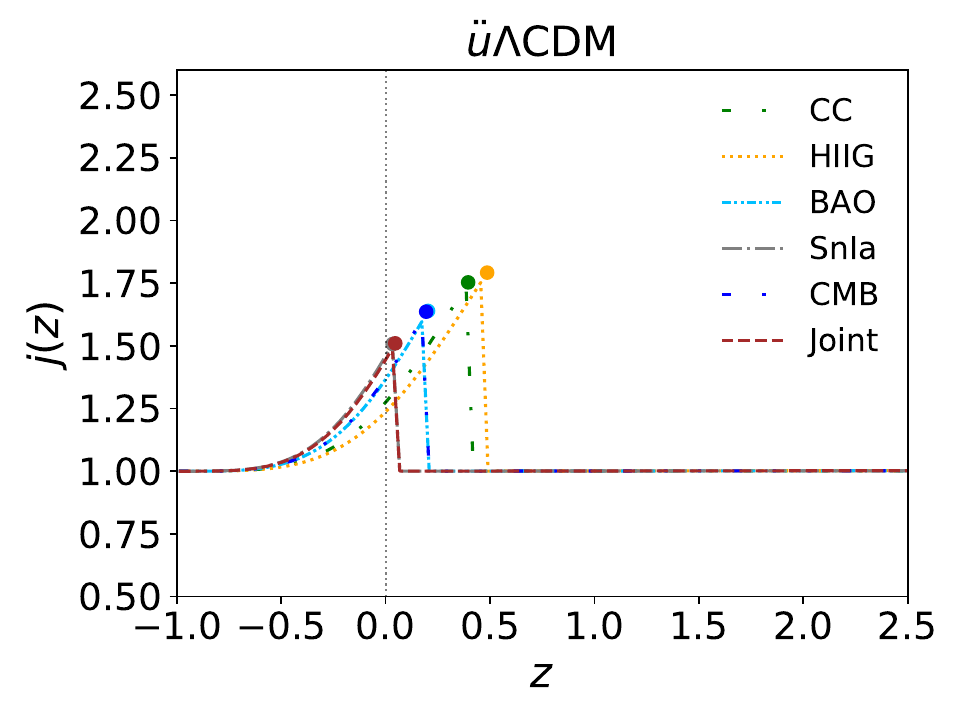} \\
\includegraphics[width=0.32\textwidth]{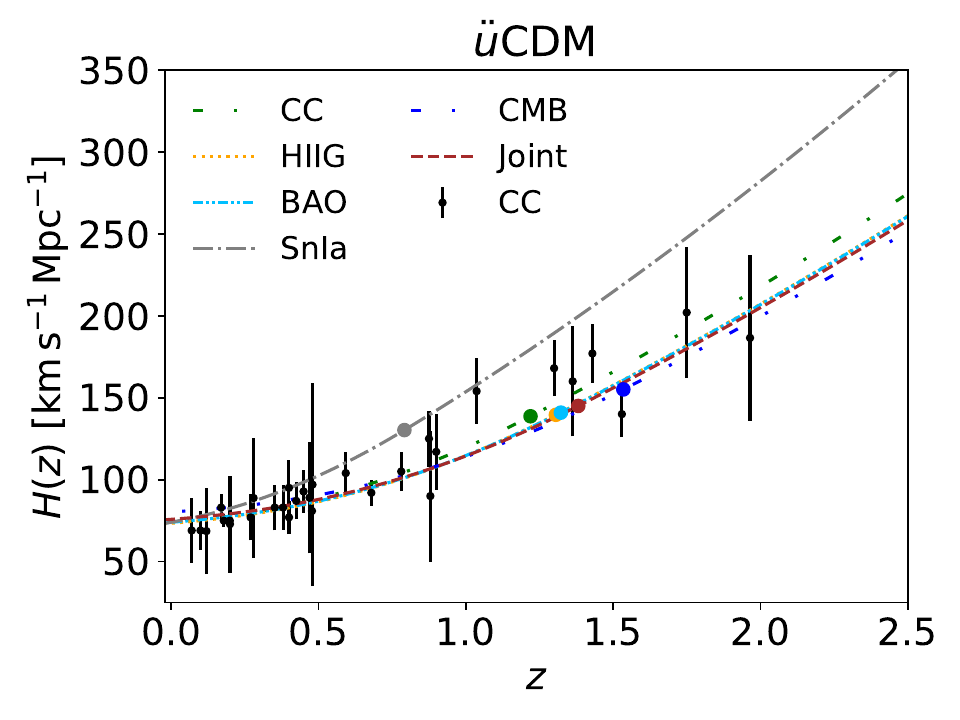}
\includegraphics[width=0.32\textwidth]{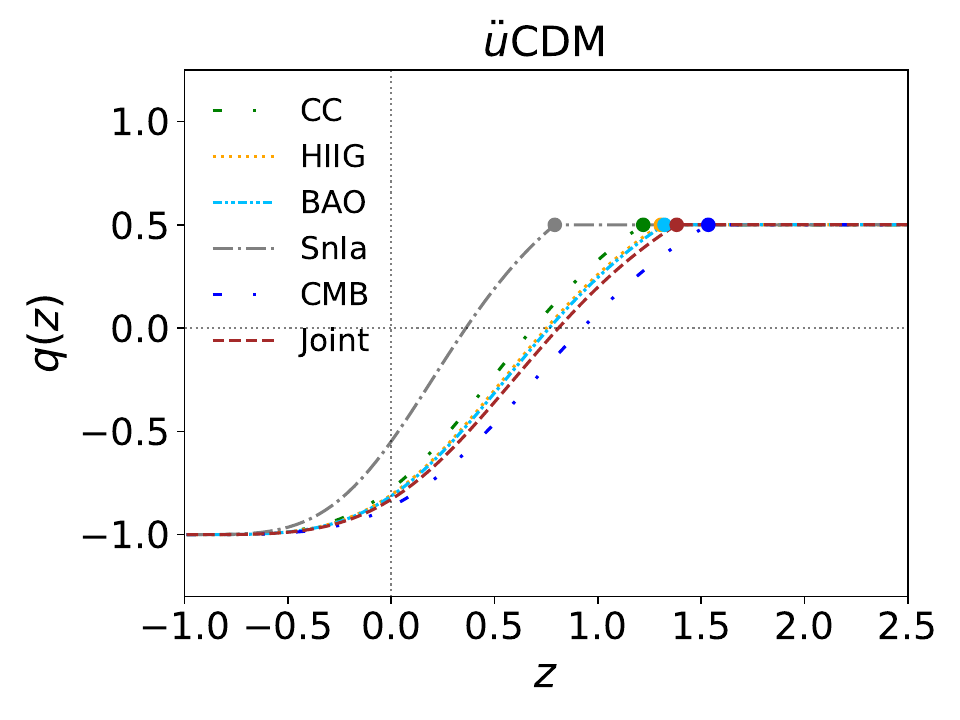}
\includegraphics[width=0.32\textwidth]{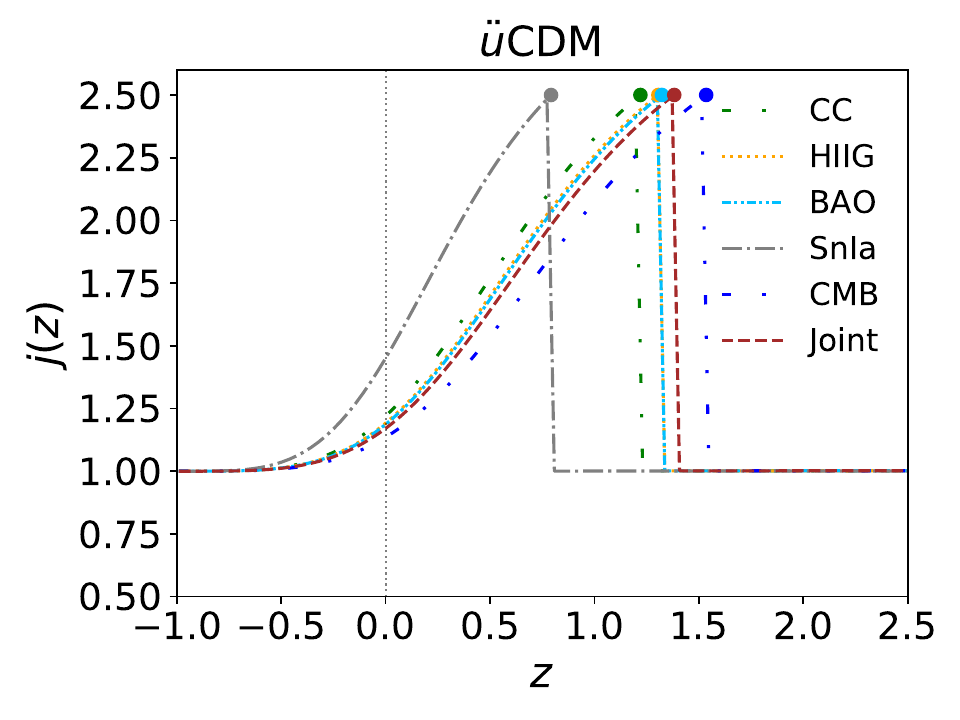}
 \caption{Reconstruction of $H(z)$ (left panel), $q(z)$ (middle panel) and $j(z)$ (right panel) for ü$\Lambda$CDM and üCDM, for each dataset. Circle marker represents the best value of $z_\oplus$ for each sample.}
\label{fig:Hz_and_qz}
\end{figure*}

\begin{figure*}
\centering
\includegraphics[width=0.41\textwidth]{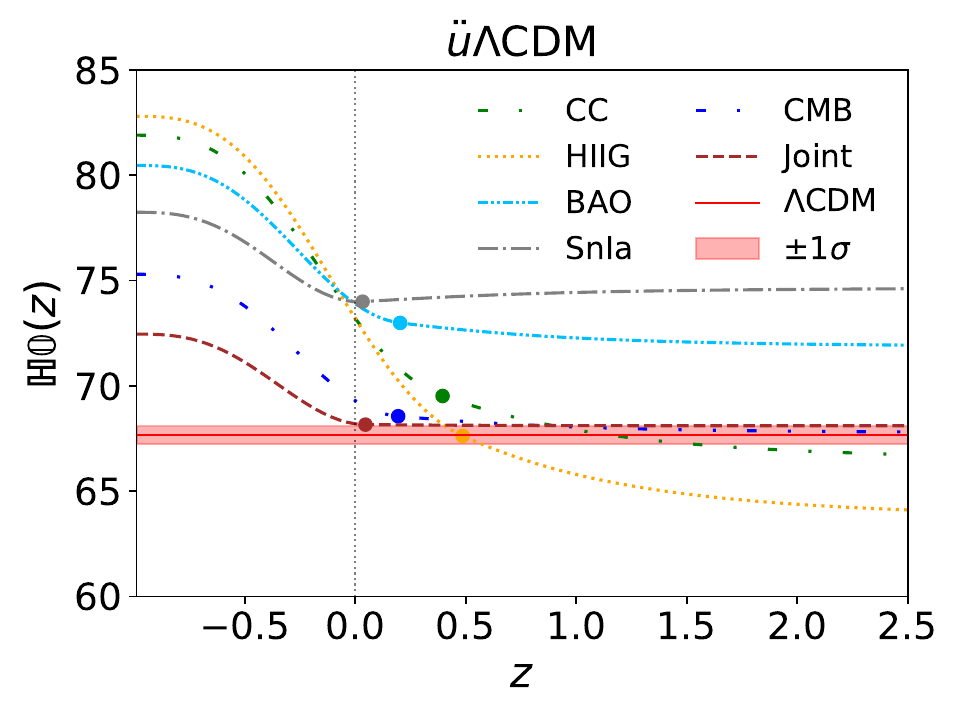}
\includegraphics[width=0.41\textwidth]{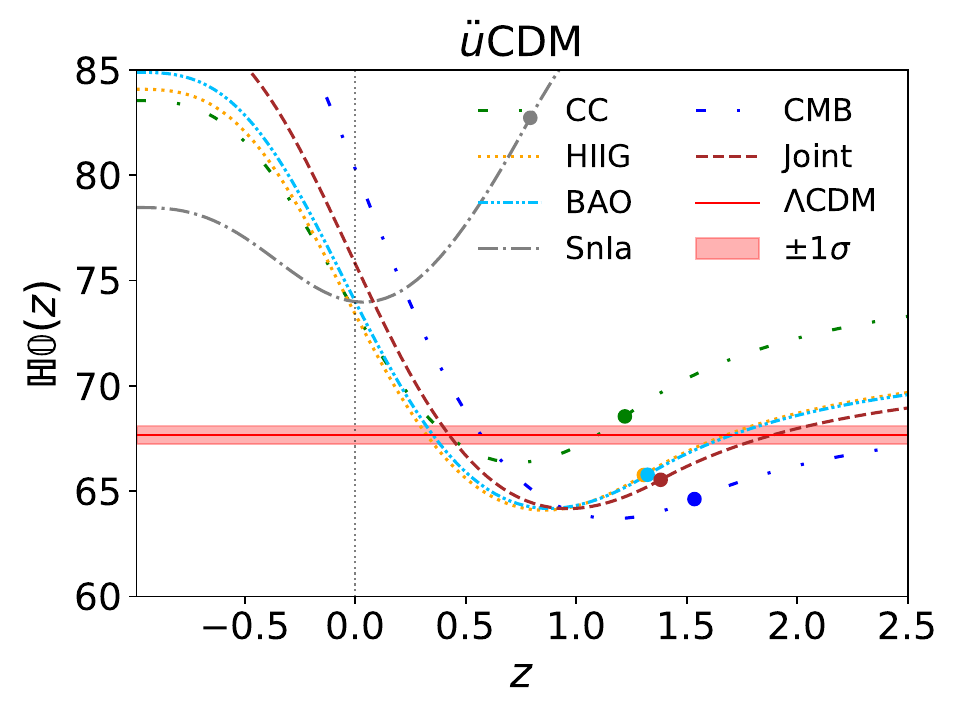}
 \caption{ The $\mathbb{H}0(z)$ diagnostic for über gravity, showing üCDM and ü$\Lambda$CDM models. Here we present the behavior of $\mathbb{H}0(z)$ under the constraints of the four data-samples and the joint analysis. The continuous red line represents the results for $\Lambda$CDM cosmology assuming $h=0.6766$ and $\Omega_{0m}=0.3111$ according to \cite{Planck:2018}.}
\label{fig:H0Diagnostic}
\end{figure*}

On the other hand, notice that the ü$\Lambda$CDM scenario produces a late acceleration, constraining the free parameter of über gravity according to the joint analysis as $z_{\oplus}\simeq 0.046$, which also coincides with the epoch of transition to über gravity. The parameter $\Omega_{0\Lambda}\simeq0.689$ marks a cosmological constant domination, however the unsolved questions about its characteristics remains. The redshift transition is compatible with $\Lambda$CDM model which is at $0.646$, as estimated from the joint analysis. The transition from GR to über gravity is perfectly observed through the jerk factor shown in Fig. \ref{fig:Hz_and_qz}, where a discontinuity in the $j(z)$ function is observed. According to the $\mathbb{H}_0$ diagnostic presented in Fig. \ref{fig:H0Diagnostic}, the tension in $H_0$ is not alleviated because it keeps the trend towards values consistent with supernovae results. 

Additionally, we explore the üCDM model which is not studied in the literature. In this case, the über component $z_{\oplus}$ acts like a cosmological constant, having the same free parameters as $\Lambda$CDM. For the joint analysis we conclude that the value of the über parameter us $z_{\oplus}\simeq 1.382$ and the transition to an accelerated Universe happens earlier in its evolution $z_{T}\simeq0.809$. From Fig. \ref{fig:Hz_and_qz}, in particular with $q(z)$ and $j(z)$, we observe that the transition to über gravity domination is more extreme in comparison with the ü$\Lambda$CDM model. Additionally to this, in this figure we also observe a trend to values that are consistent with supernova results instead of those obtained from Planck, concluding that the $H_0$ tension persist despite the über gravity presence.

\begin{table*}
\centering
\resizebox{0.7\textwidth}{!}{%
\begin{tabular}{|lccccccccc|}
\hline
Sample       &  $\chi^2_{\rm min}$     &  $h$ & $\Omega_{0\Lambda}$ & $z_\oplus$ & $z_T$ &  AICc &  $\Delta {\rm AICc}$ & BIC & $\Delta {\rm BIC}$\\ [0.9ex]
\hline
\hline
\multicolumn{10}{|c|}{\"u$\Lambda$CDM + $H_0$ SH0ES} \\ [0.9ex]
CC    & 16.24   & $0.732^{+0.013}_{-0.013}$  & $0.689^{+0.006}_{-0.006}$  & $0.395^{+0.114}_{-0.137}$  & $0.752^{+0.064}_{-0.061}$  & 23.13 & 1.06 & 26.54 & 2.03 \\ [0.9ex] 
HIIG  & 436.64  & $0.732^{+0.011}_{-0.011}$  & $0.689^{+0.006}_{-0.006}$  & $0.486^{+0.145}_{-0.167}$  & $0.803^{+0.094}_{-0.086}$  & 442.78 & 2.06 & 452.24 & 5.19 \\ [0.9ex] 
SnIa  & 2011.90 & $0.740^{+0.014}_{-0.014}$  & $0.683^{+0.005}_{-0.005}$  & $0.033^{+0.034}_{-0.023}$  & $0.628^{+0.014}_{-0.013}$  & 2017.91 & 15.50 & 2034.22 & 20.94 \\ [0.9ex] 
BAO   & 2.81    & $0.739^{+0.014}_{-0.014}$  & $0.689^{+0.006}_{-0.006}$  & $0.203^{+0.127}_{-0.133}$  & $0.675^{+0.045}_{-0.029}$  & 32.81 & 19.99 & 7.64 & 1.60 \\ [0.9ex] 
CMB   & 15.37   & $0.693^{+0.008}_{-0.007}$  & $0.687^{+0.006}_{-0.006}$  & $0.194^{+0.064}_{-0.090}$  & $0.667^{+0.016}_{-0.015}$  & 27.37 & 3.57 & 21.61 & 0.05 \\ [0.9ex] 
Joint & 2494.24 & $0.682^{+0.003}_{-0.003}$  & $0.689^{+0.004}_{-0.004}$  & $0.046^{+0.047}_{-0.032}$  & $0.646^{+0.011}_{-0.011}$  & 2500.25 & 7.00 & 2516.93 & 12.56 \\ [0.9ex]
\multicolumn{10}{|c|}{\"u$\Lambda$CDM + $H_0$ Planck} \\ [0.9ex]
CC    & 15.09   & $0.678^{+0.004}_{-0.004}$  & $0.688^{+0.005}_{-0.006}$  & $0.111^{+0.106}_{-0.077}$  & $0.654^{+0.026}_{-0.018}$  & 21.98 & 3.02 & 25.39 & 3.99 \\ [0.9ex] 
HIIG  & 441.63  & $0.679^{+0.004}_{-0.004}$  & $0.689^{+0.006}_{-0.006}$  & $0.160^{+0.146}_{-0.111}$  & $0.665^{+0.046}_{-0.024}$ & 447.77 & 2.42 & 457.23 & 5.55 \\ [0.9ex] 
SnIa  & 2011.87 & $0.676^{+0.004}_{-0.004}$  & $0.683^{+0.005}_{-0.005}$  & $0.032^{+0.034}_{-0.022}$  & $0.628^{+0.013}_{-0.013}$ & 2017.88 & 15.48 & 2034.19 & 20.92 \\ [0.9ex] 
BAO   & 2.79    & $0.677^{+0.004}_{-0.004}$  & $0.689^{+0.006}_{-0.005}$  & $0.211^{+0.124}_{-0.135}$  & $0.677^{+0.046}_{-0.030}$ & 32.79 & 19.99 & 7.62 & 1.60 \\ [0.9ex] 
CMB   & 5.40    & $0.679^{+0.003}_{-0.003}$  & $0.685^{+0.004}_{-0.004}$  & $0.055^{+0.051}_{-0.038}$  & $0.636^{+0.009}_{-0.009}$  & 17.40 & 10.50 & 11.64 & 6.98 \\ [0.9ex] 
Joint & 2473.66 & $0.678^{+0.003}_{-0.003}$  & $0.685^{+0.004}_{-0.004}$  & $0.033^{+0.035}_{-0.023}$  & $0.633^{+0.009}_{-0.009}$ & 2479.67 & 5.11 & 2496.35 & 10.67 \\ [0.9ex]
\hline
\multicolumn{10}{|c|}{\"uCDM + $H_0$ SH0ES} \\ [0.9ex]
CC    & 15.67   & $0.735^{+0.013}_{-0.013}$ & ---  & $1.220^{+0.087}_{-0.081}$  & $0.686^{+0.066}_{-0.061}$  & 20.10 & -1.97 & 22.54 & -1.97 \\ [0.9ex] 
HIIG  & 436.10  & $0.734^{+0.011}_{-0.011}$ & ---  & $1.306^{+0.125}_{-0.114}$  & $0.751^{+0.095}_{-0.087}$  & 440.17 & -0.55 & 446.50 & -0.55 \\ [0.9ex] 
SnIa  & 2017.91 & $0.740^{+0.014}_{-0.014}$ & ---  & $0.792^{+0.031}_{-0.030}$  & $0.361^{+0.023}_{-0.023}$  & 2021.92 & 19.51 & 2032.79 & 19.51 \\ [0.9ex] 
BAO   & 2.59    & $0.740^{+0.014}_{-0.014}$ & ---  & $1.323^{+0.064}_{-0.066}$  & $0.765^{+0.049}_{-0.050}$  & 12.59 & -0.23 & 5.81 & -0.23 \\ [0.9ex] 
CMB   & 26.05   & $0.803^{+0.007}_{-0.007}$ & ---  & $1.535^{+0.023}_{-0.024}$  & $0.925^{+0.018}_{-0.019}$  & 32.45 & 8.65 & 30.21 & 8.65 \\ [0.9ex] 
Joint & 2748.84 & $0.758^{+0.006}_{-0.006}$ & ---  & $1.382^{+0.020}_{-0.021}$  & $0.809^{+0.015}_{-0.016}$  & 2752.85 & 259.60 & 2763.97 & 259.60 \\ [0.9ex] 
\multicolumn{10}{|c|}{\"uCDM + $H_0$ Planck} \\ [0.9ex]
CC    & 16.81   & $0.677^{+0.004}_{-0.004}$  & --- & $1.022^{+0.067}_{-0.063}$  & $0.536^{+0.051}_{-0.048}$ & 21.24 & 2.28 & 23.68 & 2.28 \\ [0.9ex] 
HIIG  & 442.25  & $0.679^{+0.004}_{-0.004}$  & --- & $1.062^{+0.100}_{-0.094}$  & $0.566^{+0.076}_{-0.072}$ & 446.32 & 0.97 & 452.65 & 0.97 \\ [0.9ex] 
SnIa  & 2017.90 & $0.677^{+0.004}_{-0.004}$  & --- & $0.791^{+0.031}_{-0.031}$  & $0.361^{+0.024}_{-0.023}$ & 2021.91 & 19.51 & 2032.78 & 19.51 \\ [0.9ex] 
BAO   & 2.59    & $0.677^{+0.004}_{-0.004}$  & --- & $1.324^{+0.064}_{-0.065}$  & $0.766^{+0.049}_{-0.050}$ & 12.59 & -0.21 & 5.81 & -0.21 \\ [0.9ex] 
CMB   & 271.73  & $0.709^{+0.004}_{-0.004}$  & --- & $1.219^{+0.013}_{-0.014}$  & $0.685^{+0.010}_{-0.010}$ & 278.13 & 271.23 & 275.89 & 271.23 \\ [0.9ex] 
Joint & 2866.96 & $0.701^{+0.004}_{-0.004}$  & --- & $1.191^{+0.012}_{-0.013}$  & $0.665^{+0.009}_{-0.010}$ & 2870.97 & 396.41 & 2882.09 & 396.41 \\ [0.9ex]
\hline
\end{tabular}
}
\caption{ Bestfit values and their uncertainties at $1\sigma$ of the free parameters for both \"u$\Lambda$CDM and \"uCDM models using a Gaussian prior on $H_0$ on the SH0ES value and Planck value respectively. Additionally, we show values for the AICc, BIC, $\Delta{\rm AICc}\equiv {\rm AICc}-{\rm AICc}^{\Lambda \rm CDM}$ and $\Delta{\rm BIC}\equiv {\rm BIC}-{\rm BIC}^{\Lambda \rm CDM}$ for both \"u$\Lambda$CDM and \"uCDM models.}
\label{tab:bestfits}
\end{table*}



\section{Conclusions and Outlooks} \label{CO}

This paper presents an exhaustive revision of the über gravity for the ü$\Lambda$CDM and üCDM models, obtaining stringent constrictions with different cosmological data samples, in particular the recent Pantheon$+$ sample. First, we focus on our results obtained assuming a prior on $h$ provided by SHOES because these are consistent with those we obtain using a prior from Planck (see Table \ref{tab:bestfits}).
We statistically compare both models with $\Lambda$CDM by using AICc and BIC. We find that the uber cosmologies and $\Lambda$CDM are equally preferred for CC, HIIG, and BAO. When CMB data are used, \"u$\Lambda$CDM does not present evidence against but \"uCDM presents a strong evidence against. However, for both \"uber cosmologies, \"u$\Lambda$CDM and \"uCDM, the joint analysis shows the strongest evidence against them.

First, the ü$\Lambda$CDM contains one extra free parameter in comparison with the standard paradigm, maintaining the same open questions related to the understanding of the cosmological constant. Our results for the joint analysis point out that $\Omega_{0\Lambda}\simeq0.689$, consistent with the standard cosmological model, but with a subdominant value for the über parameter $z_{\oplus}\simeq0.046$. It is important to remind that the über parameter $z_{\oplus}$ is the point where the über gravity starts to dominate over the standard GR, thus, the presence of the $\Lambda$CDM delays the apparition of the über gravity as observed in Figs. \ref{fig:Hz_and_qz} for the $H(z)$, $q(z)$ and $j(z)$ parameters. The transition redshift is also in consistency with the standard cosmological model (see Table \ref{tab:bestfits}). Regarding $\mathbb{H}0(z)$ diagnostic, Fig. \ref{fig:H0Diagnostic} reveals that, according to the joint analysis, the $H_0$ value is more consistent with the supernova data than with the Planck ones at $z=0$, unable to reduce the tension under this scenario. This could be the result of $\Lambda$CDM being behind the dynamics while über gravity start its domination at $z_{\oplus}\simeq0.046$.  Additionally to this, the new sample of Pantheon+ could generate a tendency to a value greater than the one predicted by Planck. 

Furthermore, we explore the üCDM, where the über parameter plays the role of the cosmological constant, thus having the same parameters as in the standard cosmology. In this case, über parameter generates the late acceleration of the Universe and its presence start earlier, specifically at $z_{\oplus}\simeq1.382$ according to the joint analysis presented in Table \ref{tab:bestfits}. The transition also happens earlier than in the standard model, at $z_T\simeq0.809$ (see Table \ref{tab:bestfits}). The evolution of $H(z)$, $q(z)$ and $j(z)$ is presented in Fig. \ref{fig:Hz_and_qz} where we can see that the behavior is more abrupt, mainly in the $q(z)$ and $j(z)$ behavior. However, über gravity eventually mimics the cosmological constant because $j=1$ when $z=-1$. The $\mathbb{H}0(z)$ diagnostic plot presented in Fig. \ref{fig:H0Diagnostic} shows that at $z=0$ the value for $H_0$ does not coincide with the Planck result for $\Lambda$CDM, being in concordance with supernova results, which is also a sign that the über parameter acting like a cosmological constant is inadequate to alleviate the tension in $H_0$.

Finally, it is worth to point out the difference in the values of $z_{\oplus}$ for üCDM and ü$\Lambda$CDM. The discrepancy is caused mainly because in the first one the über parameter acts like a cosmological constant and not only as a transition to über gravity, which is the case for ü$\Lambda$CDM.

In summary, we observe that the über gravity is an alternative to study the late Universe acceleration by mimicking the cosmological constant behavior through the über parameter $z_{\oplus}$. However, the mystery of the Hubble tension remains unsolvable and gives values that are compatibles with supernova results. Nevertheless, we consider that a deeper exploration of the über Lagrangian for other values of $n$ is necessary in order to elucidate any extra dynamics that could help us understand the reason for the Hubble tension. Such exploration will be presented elsewhere.

\begin{acknowledgements}
We thank anonymous referee for thoughtful remarks and suggestions. G.A.C.V. thanks Dr. Teófilo Vargas Auccalla for his invaluable mentoring. C.Q. thanks the support of Universidad Nacional de San Agustin and Mg. Rolando Perca Gonz\'ales. M.A.G-A acknowledges support from c\'atedra Marcos Moshinsky (MM) and Universidad Iberoamericana for the support with the National Research System (SNI) grant. The numerical analysis was carried out by the {\it Numerical Integration for Cosmological Theory and Experiments in High-energy Astrophysics} "Nicte Ha" cluster at IBERO University, acquired through c\'atedra MM support. A.H.A. thanks to the support from Luis Aguilar, 
Alejandro de Le\'on, Carlos Flores, and Jair Garc\'ia of the Laboratorio 
Nacional de Visualizaci\'on Cient\'ifica Avanzada. V.M. acknowledges partial support from Centro de Astrofísica de Valparaíso. M.A.G.-A., A.H.A and V.M. acknowledge partial support from project ANID Vinculaci\'on Internacional FOVI220144.
\end{acknowledgements}

\bibliography{main}

\end{document}